\pdfoutput=1
\documentclass[sigconf,screen]{acmart}

\usepackage{bbm}
\usepackage{mathtools}
\usepackage[ruled, algo2e, vlined]{algorithm2e}
\usepackage{tcolorbox}
\usepackage{xcolor}

\SetCommentSty{mycommfont}

\usepackage{CJKutf8}

\allowdisplaybreaks

\AtBeginDocument{%
  \providecommand\BibTeX{{%
    \normalfont B\kern-0.5em{\scshape i\kern-0.25em b}\kern-0.8em\TeX}}}

\setcopyright{acmcopyright}
\copyrightyear{2018}
\acmYear{2018}
\acmDOI{10.1145/1122445.1122456}

\acmConference[Woodstock '18]{Woodstock '18: ACM Symposium on Neural
  Gaze Detection}{June 03--05, 2018}{Woodstock, NY}
\acmBooktitle{Woodstock '18: ACM Symposium on Neural Gaze Detection,
  June 03--05, 2018, Woodstock, NY}
\acmPrice{15.00}
\acmISBN{978-1-4503-XXXX-X/18/06}

\begin{document}

\fancyhead{}

\title{CLEAR: A Fully User-side Image Search System}

\author{Ryoma Sato}
\email{r.sato@ml.ist.i.kyoto-u.ac.jp}
\affiliation{%
  \institution{Kyoto University / RIKEN AIP}
  \city{Kyoto}
  \country{Japan}
}


\begin{abstract}
We use many search engines on the Internet in our daily lives. However, they are not perfect. Their scoring function may not model our intent or they may accept only text queries even though we want to carry out a similar image search. In such cases, we need to make a compromise: We continue to use the unsatisfactory service or leave the service. Recently, a new solution, user-side search systems, has been proposed. In this framework, each user builds their own search system that meets their preference with a user-defined scoring function and user-defined interface. Although the concept is appealing, it is still not clear if this approach is feasible in practice. In this demonstration, we show the first fully user-side image search system, CLEAR, which realizes a similar-image search engine for Flickr. The challenge is that Flickr does not provide an official similar image search engine or corresponding API. Nevertheless, CLEAR realizes it fully on a user-side. CLEAR does not use a backend server at all nor store any images or build search indices. It is in contrast to traditional search algorithms that require preparing a backend server and building a search index. Therefore, each user can easily deploy their own CLEAR engine, and the resulting service is custom-made and privacy-preserving. The online demo is available at \url{https://clear.joisino.net}. The source code is available at \url{https://github.com/joisino/clear}.
\end{abstract}



\maketitle

\section{Introduction}

A massive amount of information is uploaded on the Internet every day, and it becomes more important yet difficult to search for desired information from the flood of information. Thus, many functionalities of search engines have been called for, including multi-modal search \cite{cheng2009sketch, zha2009visual, cao2016deep, kordan2018deep} and fairness-aware systems \cite{singh2018fairness, biega2018equity, sato2022enumerating}. However, many information retrieval systems on the Internet have not adopted rich functionalities, and they usually accept simple text queries only. Even if a user of the service is unsatisfied with a search engine and is eager to enjoy additional functionalities, what he/she can do is limited. In many cases, he/she continues to use the unsatisfactory system or leaves the service.

\begin{figure}[t]
  \centering
    \includegraphics[width=\hsize]{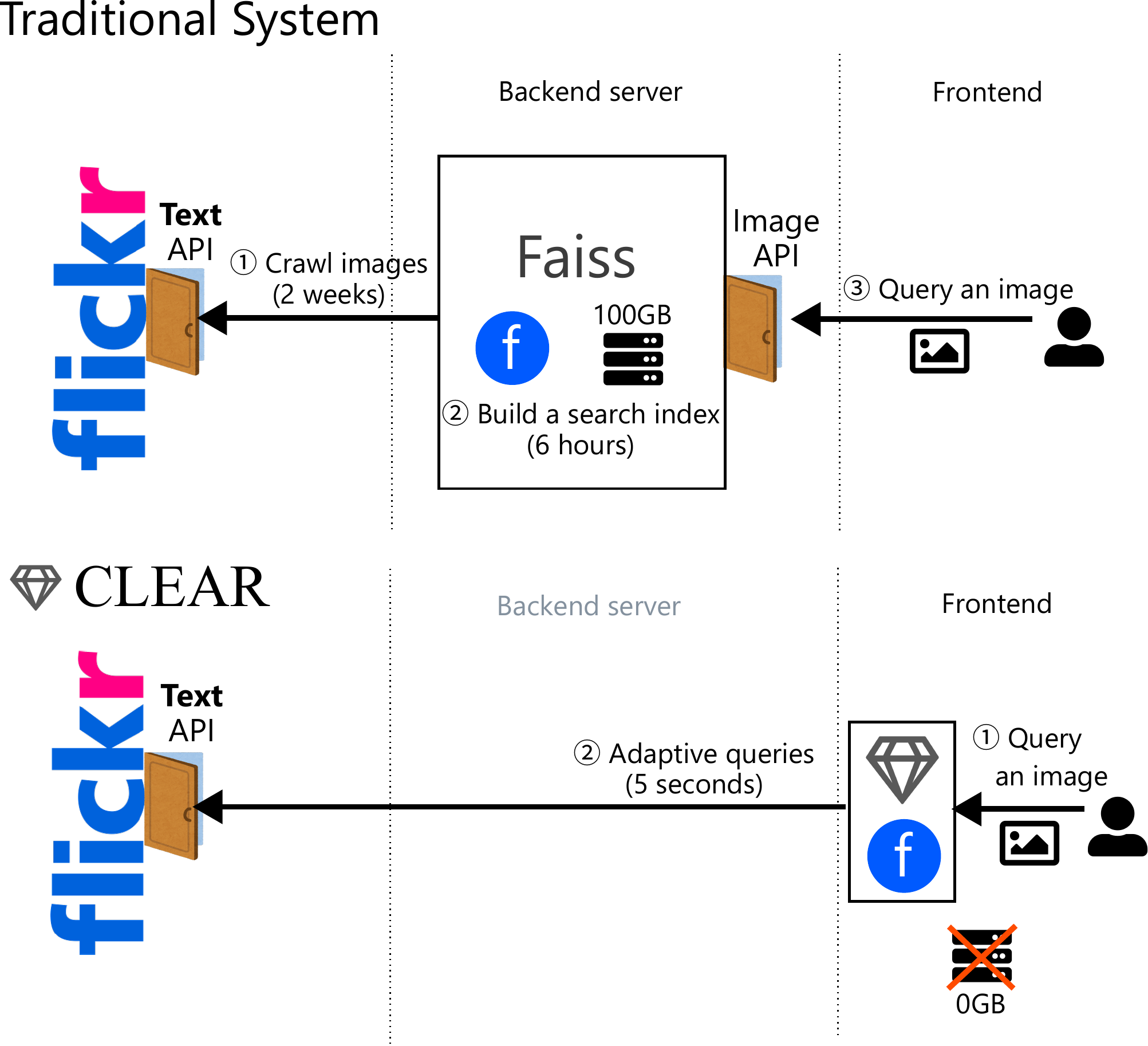}
    \vspace{-0.2in}
\caption{Flickr does not provide an official similar image search engine or corresponding API. It is not straightforward for an ordinary user to build their own search system. (Top) A traditional system would first crawl images on Flickr and build a search index. It typically takes a few weeks. (Bottom) Our proposed approach does not require initial crawling or indexing. CLEAR issues adaptive queries to the text-based image search API, which is available. It realizes an image-based search combining several effective text queries. The f mark represents the score function. In the traditional system, the score function is tightly connected to the search index. It is costly to change it. By contrast, the score function is on the frontend in CLEAR. It can be seamlessly redefined.}
 \vspace{-0.1in}
 \label{fig: illust}
\end{figure}

Let us consider Flickr\footnote{\url{https://www.flickr.com/}}, an image hosting service, as an example. Suppose we are enthusiastic users of Flickr. Although there are many amazing photos and communities on Flickr, we are not satisfied with the search engine of Flickr. Specifically, it accepts only text queries, and we cannot conduct a similar image search on Flickr. One possible yet rare solution to realize a similar image search is to join SmugMug, the owner company of Flickr, and implement the feature as an official software engineer. However, this choice is not realistic for an ordinary user.

Another possible solution is to build a search system by ourselves. We can crawl images from Flickr, build a search index by faiss \cite{johnson2019billion}, and host the search engine by ourselves. However, there are more than five billion photons on Flickr, and it typically takes a few weeks to crawl Flickr even if we selectively crawl pages with focused crawling techniques \cite{chakrabarti1999focused, mccallum2000automating, johnson2003evolving, pham2019bootstrapping}. Besides, it also requires a costly backend server to host a search engine. Therefore, this choice is neither practical for many users.

Recently, more practical user-side information retrieval algorithms have been proposed \cite{sato2022private, sato2022retrieving}. They run on a browser and do not need a backend server. However, the evaluations of existing works are conducted offline, and it is still not clear if this concept works in practice. Indeed, the official implementation of Tiara \cite{sato2022retrieving}\footnote{\url{https://github.com/joisino/tiara}}, a user-side image search algorithm, requires a few minutes for a single search, which significantly degrades user experiences and is not practical for a real-time search system. In this demonstration, we simplify and optimize Tiara and thereby show a practical user-side image search system online. \emph{This is the first practical implementation of a fully-user side search system}. We named our system CLEAR, which stands for \underbar{CL}iant-side s\underbar{EAR}ch. The features of CLEAR are as follows.
\begin{description}
\item[Lightweight.] As mentioned above and illustrated in Figure 1, traditional search systems require crawling the Web and building a search index, which takes many weeks and consumes much computational and network resources. By contrast, CLEAR does not require initial crawling or indexing, i.e., CLEAR incurs zero overhead. Each user can fork, deploy, and use CLEAR instantly. It is also easy to redefine their own scoring function. Thus, ordinary users can easily enjoy their own search systems.
\item[Fast Iteration.] When we update the feature extractor or scoring function, an ordinary system would need to rebuild the index. By contrast, as CLEAR does not use any search indices in the first place, we can seamlessly update the scoring function just by rewriting the JavaScript snippet. It accelerates the development of the desirable scoring function.
\item[Privacy-aware.] If we adopted a traditional system configuration with a backend server, we could snoop on the uploaded images in theory. By contrast, as CLEAR runs totally on the user-side, users do not have to worry about privacy issues.
\end{description}
In addition, we stress that user-side systems do not need to be used privately. Rather, each user can publish their own search engines with their own scoring function and their own interface. As CLEAR does not need a backend server, hosting a user-side system is much easier. For example, their own search system can be deployed on a static page hosting service such as Amazon S3 and GitHub Pages. We expect many characteristic user-side search systems will appear on the Internet, and it will become easy to find favorite engines. We hope CLEAR facilitates this trend.

Another use case of CLEAR is prototyping by official/unofficial developers. Even if their goal is to build a traditional system with a backend server, CLEAR accelerates the design of the score function. Besides, as CLEAR can be instantly deployed, interface designers can try and develop the system, and they can carry out user studies \emph{before} the backend engineers complete the system.

\section{Related Work}

User-side information retrieval systems \cite{sato2022private, sato2022retrieving} enable each user to build their own system, whereas traditional systems are developed by service developers. For example, private recommender systems \cite{sato2022private} turn recommendation results into fair and/or diverse ones even if the official recommender systems are not fair or diverse. Tiara \cite{sato2022retrieving} is the most relevant work to this demonstration. It realizes image retrieval based on user-defined score functions. The critical difference between Tiara and this demonstration is that the original paper of Tiara conducted only batch evaluations. Indeed, the official implementation of Tiara requires a few minutes for a single search on Flickr, which is too slow for real-time demonstration. We optimize the system and realize a real-time user-side search system.

A steerable system \cite{green2009generating, balog2019transparent} also allows users to customize the system. However, the critical difference with user-side systems is that steerable systems are implemented by official developers, and users cannot enjoy this feature if the official system is not steerable in the first place. By contrast, user-side systems turn ordinary systems into steerable ones on the user-side.

Another relevant realm is focused crawling \cite{chakrabarti1999focused, baezayates2005crawling, barbosa2007adaptive, guan2008guide, meusel2014focused}. In contrast to exhaustive crawling, focused crawling aims to retrieve only relevant information. Although these techniques accelerate crawling, they still require several hours to several weeks. Therefore, most users cannot afford to adopt such systems due to time, network, or computational resources. In stark contrast, our system does not require initial crawling at all.

\section{CLEAR}

\begin{figure}[t]
  \centering
    \includegraphics[width=0.9\hsize]{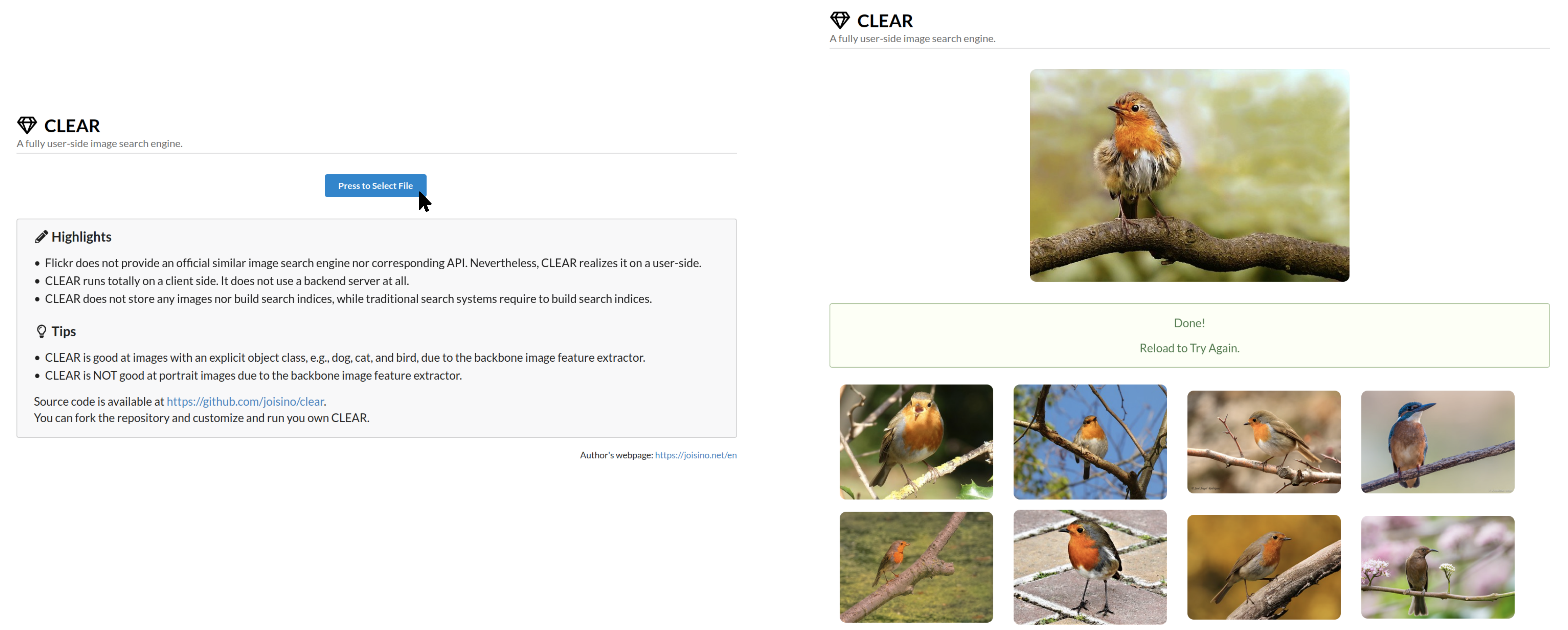}
 \vspace{-0.1in}
\caption{Interface of CLEAR. Upload an image, and then CLEAR retrieves similar images from Flickr. The functionality of CLEAR is ordinary. The highlight lies rather in how it is realized and how easy deployment is. (Left) User interface. (Right) Search results.}
 \vspace{-0.3in}
 \label{fig: interface}
\end{figure}

\subsection{User Interface}

The interface of CLEAR is simple (Figure \ref{fig: interface}). A user clicks the blue button and uploads an image; then CLEAR retrieves similar images from Flickr. The functionality is an ordinal similar image search. We stress that the highlight of CLEAR lies rather in how it is realized. It is remarkable that CLEAR works similarly to an ordinary system with a backend server.

\subsection{Algorithm}

\begin{figure}[t]
  \centering
    \includegraphics[width=0.8\hsize]{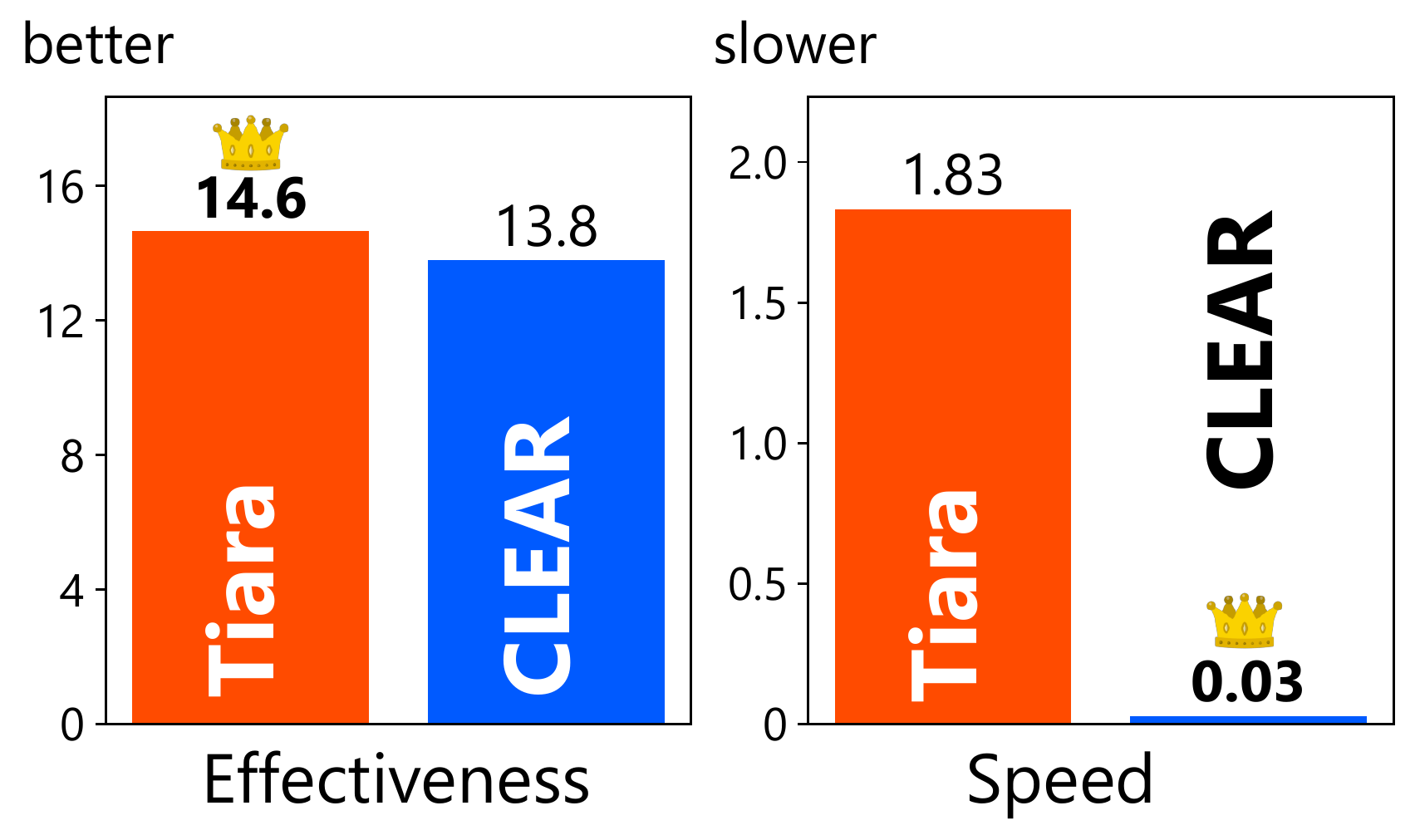}
\caption{The greedy algorithm is fast while the degradation of performance is slight. The better one is highlighted with bold style and a crown mark on each panel. (Left) Each value is the average score of the retrieved images. The higher, the better the retrieved images are. This shows that the greedy algorithm slightly degrades the performance. We confirmed that the retrieved images by both methods were visually comparable. (Right) The average running time in seconds. The greedy method is $60$ times more efficient.}
 \vspace{-0.1in}
 \label{fig: experiments}
\end{figure}

\noindent \textbf{Preliminary: Tiara.} We first introduce Tiara \cite{sato2022retrieving}, on which CLEAR is based. Tiara assumes that we have a black-box score function $f$ that takes an image as input and outputs a real-valued score. This score function is typically realized by an off-the-shelf feature extractor such as ResNet and MobileNet. The aim of Tiara is to find images with high scores from an external image database, such as Flickr. The critical assumption is that we do not have direct access to the external database, but we can access it via a tag-based search API only. The core idea of Tiara is that it regards a tag as an arm and the score function $f$ as a reward and formulates this problem as a multi-armed bandit algorithm. Tiara uses the linear bandit algorithm \cite{li2010contextual} with a UCB-style acquisition function \cite{auer2002using}. Tiara utilizes the GloVe word embeddings \cite{pennington2014glove} for tag features to cope with the cold-start problem. Tiara realized an effective similar image search on Flickr with a similarity function $f$.

\vspace{0.1in}
\noindent \textbf{CLEAR.} We design the algorithm for CLEAR based on Tiara. Although the use of the bandit algorithm adopted in Tiara strikes a good tradeoff between effectiveness and efficiency, it consumes too much time, e.g., a few minutes, for a real-time search. Besides, the GloVe embeddings consume about $1$ GB, which makes it difficult to download the program on the client-side. We find that the greedy algorithm that always draws the current best arm is still effective while it's much more efficient. The greedy algorithm corresponds to setting $\lambda = 0.0$ in Tiara, i.e., no exploration. The greedy algorithm is efficient because (i) it does not need to compute the matrix inverse for computing the UCB score, (ii) it does not consume the query budget for exploration, and (iii) it does not require downloading word embeddings. Indeed, exploration may lead to better search results in the long run, but the real-time search in CLEAR runs only several iterations, and thus the greedy approach is sufficient. Besides, we found that the response of an API call was stochastic, and therefore, the greedy algorithm also did some exploration rather than querying only one arm.

To quantify the effect of using the greedy algorithm, we conducted an offline evaluation using the Open Image Dataset \cite{OpenImages}. The experimental setup is the same as in Tiara \cite{sato2022retrieving}. Specifically, we use a subset of the Open Image Dataset with $100\,000$ images as a virtual image repository and use the labels of an image as a set of tags. We use the same score function as Tiara, i.e., the logit of ResNet. We set the query budget as $100$ in this experiment. Figure \ref{fig: experiments} shows that the greedy algorithm slightly degrades the performance while it is $60$ times more efficient. Tiara is effective when there is no tight time constraint. As we call for real-time inference, we adopt the greedy algorithm in the demonstration.

\subsection{Implementation}

CLEAR is implemented with the React framework. It adaptively calls \texttt{flickr.photos.search} API via the Axios library. This API takes a text as input and returns a set of images and their attached tags. CLEAR manages the average score of each tag and queries the tag with the highest average score iteratively. CLEAR shows the returned images in the order of scores. CLEAR's score function is \begin{align}
f(x) \stackrel{\text{def}}{=} \exp(g(x)^\top g(s) / 1000),
\end{align} where $g$ is the MobileNetv2 \cite{sandler2018mobilenetv2} feature extractor and $s$ is the source image uploaded by the user. The higher the inner product similarity to the source image, the higher this score is. The feature extractor is implemented with the TensorFlow.js library \cite{smilkov2019tensorflow}. To improve the effectiveness and user experience, CLEAR adopts the following strategies.
\begin{description}
\item[Initial queries.] At first, CLEAR does not have any candidate tags. To generate initial queries, CLEAR first obtains class names of the input image using the MobileNetv2 classifier and uses top-$k$ class names for initial queries. This is in contrast to Tiara, which uses a fixed set of initial tags.
\item[Parallelization.] As an API call is the bottleneck of wall-time consumption, CLEAR issues several API calls simultaneously and aggregates results. Specifically, CLEAR selects top-$10$ tags and issues $10$ queries at once. The score evaluation is also parallelized. This is in contrast to Tiara, which issues a single query at once.
\item[Date Specification.] The \texttt{flickr.photos.search} API can specify the dates of images, and the default setting is the latest images. We found that the default setting was not effective and reduced diversity. For example, we upload a cat image, and ``cat'' and ``calico cat'' tags are the identified best tags. CLEAR queries both ``cat'' and ``calico cat'' tags, and the set of returned images would largely overlap each other. To overcome this issue, we randomly set the dates for each tag so that the returned images do not overlap one another.
\item[Real-time display.] Although CLEAR is efficient, it still takes tens of seconds to complete a single search, which degrades the user experience. To mitigate this, CLEAR shows the intermediate results, and thereby, users can see the first results in a few seconds. This improves the user experience much.
\end{description}

We also note that the search results of CLEAR are stochastic. First, the Flickr API is stochastic because many images are being uploaded and some images are being deleted. Second, CLEAR's algorithm itself is stochastic in specifying the dates and selecting tags. As the former one is difficult to remove, we keep the entire system stochastic. The demerit of this choice would be the lack of reproducibility. However, the merit is that a user can try the same query many times and obtain fresh information every time. This improves the chance of serendipitous findings. 

\subsection{Bespoke System}

The source code of CLEAR is available at \url{https://github.com/joisino/clear}. An interested user can fork the repository and build their own system. As mentioned earlier, CLEAR is easy to deploy and host because it is a fully-user side system. The score function is defined in \texttt{src/score.js}. A user can rewrite the function and instantly tries their own scoring function. A small change in the score function may change the behavior of the system, and a user may be able to find a search engine that fits his/her preference.
\begin{itemize}
    \item \texttt{getFeature} function defined in \texttt{src/score.js} computes feature vectors for both the source image and retrieved images. The embedding layer is defined in \texttt{embeddingName}. This demonstration uses \\ \texttt{module\_apply\_default/MobilenetV2/\\expanded\_conv\_16/project/BatchNorm/FusedBatchNorm} layer. One can try other layers, e.g., \\ \texttt{module\_apply\_default/MobilenetV2/Logits/AvgPool}.
    \item \texttt{embs2score} function defined in \texttt{src/score.js} computes scores. The higher the better. This demonstration uses \\ \texttt{Math.exp(emb1.mul(emb2).sum().dataSync()[0] / 1000)}. One can try other functions, e.g., the Gaussian kernel \\ \texttt{Math.exp(- emb1.squaredDifference(emb2).sum() \\ .dataSync()[0] / 1000)}.
\end{itemize}
Although we focused on a similar image search system in this demonstration, the score function needs not to measure similarity. If one trains a neural network using his/her favorite images so that preferable images have high scores and uses it as the score function, then the resulting system searches for preferable images. If a fairness-aware score function is used, a fairness-aware search system is realized. Another interesting setting is to use a black-box (possibly buggy) neural network as the score function as proposed in Tiara \cite{sato2022retrieving}. CLEAR finds a set of images that activate the neural network. Such instances visually show what the neural network at hand represents \cite{simonyan2014deep, nguyen2016synthesizing, yuan2020xgnn}. Traditional interpretation methods rely on a fixed set of datasets such as ImageNet, but CLEAR can retrieve images from Flickr, which hosts as many as five billion images. In sum, the creativity for the design of the score function is open to each user.

As CLEAR does not rely on any backend servers or search indices, one can seamlessly use the system after one changes the score function. One can also change the search target from Flickr to other services by writing a wrapper in \texttt{src/flickr.js}.

Note also that the online demo we provide limits the number of queries because many users may use it simultaneously. A bespoke system allows using more query budgets if needed.

\section{Demonstration}

\begin{figure}[t]
  \centering
    \includegraphics[width=\hsize]{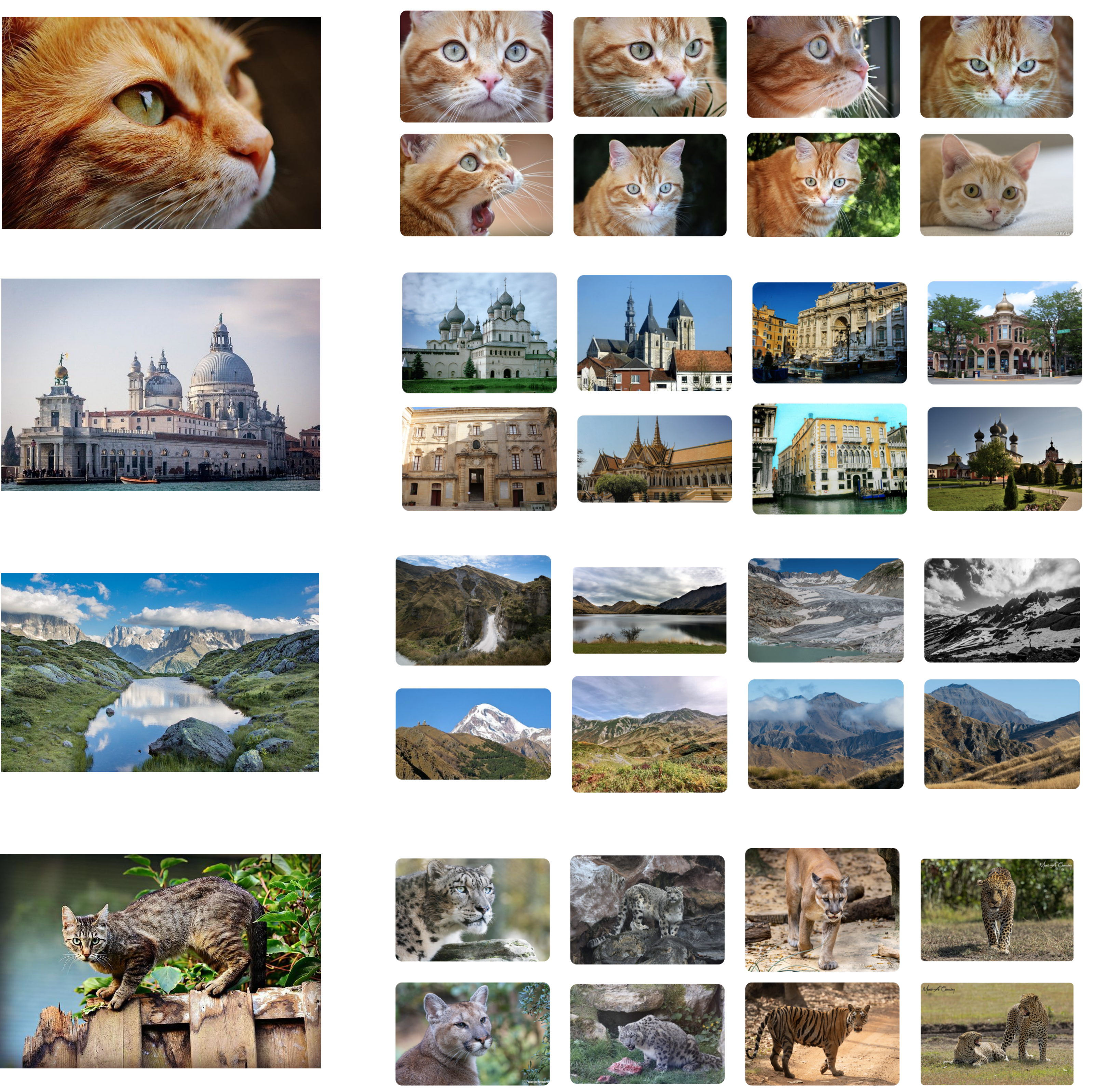}
\caption{Results. (Left) Source images. (Right) Top-$8$ retrieved images by CLEAR. (Bottom) A failure example. Although the input image is a cat, the retrieved images are snow leopards.}
 \vspace{-0.1in}
 \label{fig: result}
\end{figure}

The online demo is available at \url{https://clear.joisino.net}. Users can upload images, and the system retrieves similar images from Flickr. The functionality is simple. We stress again that Flickr does not provide an official similar image search system, and therefore users could not enjoy this feature so far. The highlight is that although the author is not employed by Flickr nor has privileged access to Flicker's server, CLEAR realizes the new feature of Flickr on a user-side.

\vspace{0.1in}
\noindent \textbf{Results.} Figure \ref{fig: result} shows results of the demo program. It can be observed that CLEAR retrieves visually similar images to the source images. The bottom row shows a failure example. Although the input image is a cat, CLEAR confused this image with a snow leopard. Once CLEAR becomes confident to some extent, it is difficult to escape from the failure mode because it adopts the greedy algorithm without explicit exploration. Improving stability while keeping efficiency is an important future direction.

\section{Conclusion}

This demonstration provides the first practical user-side image search system, CLEAR. The proposed system is implemented on the client-side and retrieves similar images from Flickr, although Flickr does not provide an official similar image search engine or corresponding API. CLEAR adaptively generates effective text queries and realizes a similar image search. CLEAR does not require the initial crawling, indexing, or any backend servers. Therefore, an ordinary user can deploy their own search engine easily.

\begin{acks}
This work was supported by JSPS KAKENHI GrantNumber 21J22490. 
\end{acks}


\bibliographystyle{plainnat}
\bibliography{sample-base}

\begin{thebibliography}{29}
\providecommand{\natexlab}[1]{#1}
\providecommand{\url}[1]{\texttt{#1}}
\expandafter\ifx\csname urlstyle\endcsname\relax
  \providecommand{\doi}[1]{doi: #1}\else
  \providecommand{\doi}{doi: \begingroup \urlstyle{rm}\Url}\fi

\bibitem[Auer(2002)]{auer2002using}
Peter Auer.
\newblock Using confidence bounds for exploitation-exploration trade-offs.
\newblock \emph{J. Mach. Learn. Res.}, 3:\penalty0 397--422, 2002.

\bibitem[Baeza{-}Yates et~al.(2005)Baeza{-}Yates, Castillo, Mar{\'{\i}}n, and
  Rodr{\'{\i}}guez]{baezayates2005crawling}
Ricardo~A. Baeza{-}Yates, Carlos Castillo, Mauricio Mar{\'{\i}}n, and M.~Andrea
  Rodr{\'{\i}}guez.
\newblock Crawling a country: better strategies than breadth-first for web page
  ordering.
\newblock In \emph{Proceedings of the 2005 World Wide Web Conference, {WWW}},
  pages 864--872, 2005.

\bibitem[Balog et~al.(2019)Balog, Radlinski, and
  Arakelyan]{balog2019transparent}
Krisztian Balog, Filip Radlinski, and Shushan Arakelyan.
\newblock Transparent, scrutable and explainable user models for personalized
  recommendation.
\newblock In \emph{Proceedings of the 42nd International {ACM} {SIGIR}
  Conference on Research and Development in Information Retrieval, {SIGIR}},
  pages 265--274, 2019.

\bibitem[Barbosa and Freire(2007)]{barbosa2007adaptive}
Luciano Barbosa and Juliana Freire.
\newblock An adaptive crawler for locating hidden-web entry points.
\newblock In \emph{Proceedings of the 2007 World Wide Web Conference, {WWW}},
  pages 441--450, 2007.

\bibitem[Biega et~al.(2018)Biega, Gummadi, and Weikum]{biega2018equity}
Asia~J. Biega, Krishna~P. Gummadi, and Gerhard Weikum.
\newblock Equity of attention: Amortizing individual fairness in rankings.
\newblock In \emph{Proceedings of the 41st International {ACM} {SIGIR}
  Conference on Research and Development in Information Retrieval, {SIGIR}},
  pages 405--414. {ACM}, 2018.

\bibitem[Cao et~al.(2016)Cao, Long, Wang, Yang, and Yu]{cao2016deep}
Yue Cao, Mingsheng Long, Jianmin Wang, Qiang Yang, and Philip~S. Yu.
\newblock Deep visual-semantic hashing for cross-modal retrieval.
\newblock In \emph{Proceedings of the 22nd {ACM} {SIGKDD} International
  Conference on Knowledge Discovery and Data Mining, {KDD}}, pages 1445--1454,
  2016.

\bibitem[Chakrabarti et~al.(1999)Chakrabarti, van~den Berg, and
  Dom]{chakrabarti1999focused}
Soumen Chakrabarti, Martin van~den Berg, and Byron Dom.
\newblock Focused crawling: {A} new approach to topic-specific web resource
  discovery.
\newblock \emph{Comput. Networks}, 31\penalty0 (11-16):\penalty0 1623--1640,
  1999.

\bibitem[Chen et~al.(2009)Chen, Cheng, Tan, Shamir, and Hu]{cheng2009sketch}
Tao Chen, Ming{-}Ming Cheng, Ping Tan, Ariel Shamir, and Shi{-}Min Hu.
\newblock Sketch2photo: internet image montage.
\newblock \emph{{ACM} Trans. Graph.}, 28\penalty0 (5):\penalty0 124, 2009.

\bibitem[Green et~al.(2009)Green, Lamere, Alexander, Maillet, Kirk, Holt,
  Bourque, and Mak]{green2009generating}
Stephen~J. Green, Paul Lamere, Jeffrey Alexander, Fran{\c{c}}ois Maillet,
  Susanna Kirk, Jessica Holt, Jackie Bourque, and Xiao{-}Wen Mak.
\newblock Generating transparent, steerable recommendations from textual
  descriptions of items.
\newblock In \emph{Proceedings of the 2009 {ACM} Conference on Recommender
  Systems, {RecSys}}, pages 281--284, 2009.

\bibitem[Guan et~al.(2008)Guan, Wang, Chen, Bu, and Wang]{guan2008guide}
Ziyu Guan, Can Wang, Chun Chen, Jiajun Bu, and Junfeng Wang.
\newblock Guide focused crawler efficiently and effectively using on-line
  topical importance estimation.
\newblock In \emph{Proceedings of the 31st International {ACM} {SIGIR}
  Conference on Research and Development in Information Retrieval, {SIGIR}},
  pages 757--758, 2008.

\bibitem[Johnson et~al.(2021)Johnson, Douze, and
  J{\'{e}}gou]{johnson2019billion}
Jeff Johnson, Matthijs Douze, and Herv{\'{e}} J{\'{e}}gou.
\newblock Billion-scale similarity search with gpus.
\newblock \emph{{IEEE} Trans. Big Data}, 7\penalty0 (3):\penalty0 535--547,
  2021.

\bibitem[Johnson et~al.(2003)Johnson, Tsioutsiouliklis, and
  Giles]{johnson2003evolving}
Judy Johnson, Kostas Tsioutsiouliklis, and C.~Lee Giles.
\newblock Evolving strategies for focused web crawling.
\newblock In \emph{Proceedings of the Twentieth International Conference on
  Machine Learning, {ICML}}, pages 298--305, 2003.

\bibitem[Kordan and Kotov(2018)]{kordan2018deep}
Saeid~Balaneshin Kordan and Alexander Kotov.
\newblock Deep neural architecture for multi-modal retrieval based on joint
  embedding space for text and images.
\newblock In \emph{Proceedings of the 11th {ACM} International Conference on
  Web Search and Data Mining, {WSDM}}, pages 28--36, 2018.

\bibitem[Kuznetsova et~al.(2020)Kuznetsova, Rom, Alldrin, Uijlings, Krasin,
  Pont-Tuset, Kamali, Popov, Malloci, Kolesnikov, Duerig, and
  Ferrari]{OpenImages}
Alina Kuznetsova, Hassan Rom, Neil Alldrin, Jasper Uijlings, Ivan Krasin, Jordi
  Pont-Tuset, Shahab Kamali, Stefan Popov, Matteo Malloci, Alexander
  Kolesnikov, Tom Duerig, and Vittorio Ferrari.
\newblock The open images dataset v4: Unified image classification, object
  detection, and visual relationship detection at scale.
\newblock \emph{Int. J. Comput. Vis.}, 128\penalty0 (7):\penalty0 1956--1981,
  2020.

\bibitem[Li et~al.(2010)Li, Chu, Langford, and Schapire]{li2010contextual}
Lihong Li, Wei Chu, John Langford, and Robert~E. Schapire.
\newblock A contextual-bandit approach to personalized news article
  recommendation.
\newblock In \emph{Proceedings of the 2010 World Wide Web Conference, {WWW}},
  pages 661--670, 2010.

\bibitem[McCallum et~al.(2000)McCallum, Nigam, Rennie, and
  Seymore]{mccallum2000automating}
Andrew McCallum, Kamal Nigam, Jason Rennie, and Kristie Seymore.
\newblock Automating the construction of internet portals with machine
  learning.
\newblock \emph{Inf. Retr.}, 3\penalty0 (2):\penalty0 127--163, 2000.

\bibitem[Meusel et~al.(2014)Meusel, Mika, and Blanco]{meusel2014focused}
Robert Meusel, Peter Mika, and Roi Blanco.
\newblock Focused crawling for structured data.
\newblock In \emph{Proceedings of the 23rd {ACM} International Conference on
  Conference on Information and Knowledge Management, {CIKM}}, pages
  1039--1048, 2014.

\bibitem[Nguyen et~al.(2016)Nguyen, Dosovitskiy, Yosinski, Brox, and
  Clune]{nguyen2016synthesizing}
Anh~Mai Nguyen, Alexey Dosovitskiy, Jason Yosinski, Thomas Brox, and Jeff
  Clune.
\newblock Synthesizing the preferred inputs for neurons in neural networks via
  deep generator networks.
\newblock In \emph{Advances in Neural Information Processing Systems 29: Annual
  Conference on Neural Information Processing Systems 2016, {NeurIPS}}, pages
  3387--3395, 2016.

\bibitem[Pennington et~al.(2014)Pennington, Socher, and
  Manning]{pennington2014glove}
Jeffrey Pennington, Richard Socher, and Christopher~D. Manning.
\newblock Glove: Global vectors for word representation.
\newblock In \emph{Proceedings of the 2014 Conference on Empirical Methods in
  Natural Language Processing, {EMNLP}}, pages 1532--1543, 2014.

\bibitem[Pham et~al.(2019)Pham, Santos, and Freire]{pham2019bootstrapping}
Kien Pham, A{\'{e}}cio S.~R. Santos, and Juliana Freire.
\newblock Bootstrapping domain-specific content discovery on the web.
\newblock In \emph{Proceedings of the 2019 World Wide Web Conference, {WWW}},
  pages 1476--1486, 2019.

\bibitem[Sandler et~al.(2018)Sandler, Howard, Zhu, Zhmoginov, and
  Chen]{sandler2018mobilenetv2}
Mark Sandler, Andrew~G. Howard, Menglong Zhu, Andrey Zhmoginov, and
  Liang{-}Chieh Chen.
\newblock Mobilenetv2: Inverted residuals and linear bottlenecks.
\newblock In \emph{Proceedings of the 2018 {IEEE} Conference on Computer Vision
  and Pattern Recognition, {CVPR}}, pages 4510--4520, 2018.

\bibitem[Sato(2022{\natexlab{a}})]{sato2022enumerating}
Ryoma Sato.
\newblock Enumerating fair packages for group recommendations.
\newblock In \emph{Proceedings of the 15th {ACM} International Conference on
  Web Search and Data Mining, {WSDM}}, pages 870--878. {ACM},
  2022{\natexlab{a}}.

\bibitem[Sato(2022{\natexlab{b}})]{sato2022private}
Ryoma Sato.
\newblock Private recommender systems: How can users build their own fair
  recommender systems without log data?
\newblock In \emph{Proceedings of the 2022 {SIAM} International Conference on
  Data Mining, {SDM}}, 2022{\natexlab{b}}.

\bibitem[Sato(2022{\natexlab{c}})]{sato2022retrieving}
Ryoma Sato.
\newblock Retrieving black-box optimal images from external databases.
\newblock In \emph{Proceedings of the 15th {ACM} International Conference on
  Web Search and Data Mining, {WSDM}}, pages 879--887. {ACM},
  2022{\natexlab{c}}.

\bibitem[Simonyan et~al.(2014)Simonyan, Vedaldi, and
  Zisserman]{simonyan2014deep}
Karen Simonyan, Andrea Vedaldi, and Andrew Zisserman.
\newblock Deep inside convolutional networks: Visualising image classification
  models and saliency maps.
\newblock In \emph{Proceedings of the 2nd International Conference on Learning
  Representations, {ICLR}, Workshop Track Proceedings}, 2014.

\bibitem[Singh and Joachims(2018)]{singh2018fairness}
Ashudeep Singh and Thorsten Joachims.
\newblock Fairness of exposure in rankings.
\newblock In \emph{Proceedings of the 24th {ACM} {SIGKDD} International
  Conference on Knowledge Discovery and Data Mining, {KDD}}, pages 2219--2228,
  2018.

\bibitem[Smilkov et~al.(2019)Smilkov, Thorat, Assogba, Yuan, Kreeger, Yu,
  Zhang, Cai, Nielsen, Soergel, Bileschi, Terry, Nicholson, Gupta, Sirajuddin,
  Sculley, Monga, Corrado, Vi{\'{e}}gas, and Wattenberg]{smilkov2019tensorflow}
Daniel Smilkov, Nikhil Thorat, Yannick Assogba, Ann Yuan, Nick Kreeger, Ping
  Yu, Kangyi Zhang, Shanqing Cai, Eric Nielsen, David Soergel, Stan Bileschi,
  Michael Terry, Charles Nicholson, Sandeep~N. Gupta, Sarah Sirajuddin,
  D.~Sculley, Rajat Monga, Greg Corrado, Fernanda~B. Vi{\'{e}}gas, and Martin
  Wattenberg.
\newblock Tensorflow.js: Machine learning for the web and beyond.
\newblock In \emph{Proceedings of Machine Learning and Systems 2019, {MLSys}}.
  mlsys.org, 2019.

\bibitem[Yuan et~al.(2020)Yuan, Tang, Hu, and Ji]{yuan2020xgnn}
Hao Yuan, Jiliang Tang, Xia Hu, and Shuiwang Ji.
\newblock {XGNN:} towards model-level explanations of graph neural networks.
\newblock In \emph{Proceedings of the 26th {ACM} {SIGKDD} International
  Conference on Knowledge Discovery and Data Mining, {KDD}}, pages 430--438,
  2020.

\bibitem[Zha et~al.(2009)Zha, Yang, Mei, Wang, and Wang]{zha2009visual}
Zheng{-}Jun Zha, Linjun Yang, Tao Mei, Meng Wang, and Zengfu Wang.
\newblock Visual query suggestion.
\newblock In \emph{Proceedings of the 17th International Conference on
  Multimedia, {MM}}, pages 15--24. {ACM}, 2009.

\end{thebibliography}










\end{document}